\documentclass[prd,aps]{revtex4}
\usepackage{amssymb,amsmath,amsfonts,amsbsy,epsfig}
\usepackage{slashed}
\usepackage{xcolor}
\newcommand{\be}{\begin{equation}}
\newcommand{\ee}{\end{equation}}
\newcommand{\bea}{\begin{eqnarray}}
\newcommand{\eea}{\end{eqnarray}}
\newcommand{\nn}{\nonumber}

\def\L{{\mathcal L}}

\begin{document}

\title{Meson spectrum in $QCD_2$ revisited}
\date{\today}

\author{
Jorge Alfaro$^{a}$, and Alex Soto$^{a}$
}

\affiliation{
$^{a}$Instituto de F\'{i}sica, Pontificia Universidad de Cat\'olica de Chile, \mbox{Av. Vicu\~na Mackenna 4860, Santiago, Chile}
}

\begin{abstract} 

Recently it has been shown that in two dimensions is possible to add new Lorentz invariant terms built with fractions containing the null vector $n=(1,1)$. In this work, we have computed the meson spectrum following the 't Hooft model in $QCD_2$ incorporating these new kinds of terms. We found these new terms does not affect the meson spectrum. We have computed the 't Hooft model with a new regulator. We have introduced a gluon mass and we have recovered the 't Hooft result when this parameter is set to zero.

\end{abstract}

\maketitle 

\section{Introduction}

After the discovery of asymptotic freedom in Quantum Chromodynamics (QCD), this theory has been well accepted as a good description of the strong interactions. In spite of this fact, our understanding of the non-perturbative limit is not completely clear. To gain a better comprehension of some problems like confinement and bound-state spectrum, among others, is usual to work in lower dimensions. In this way, $QCD_2$, whose study was pioneered by 't Hooft\cite{tHooft:1973alw,tHooft:1974pnl}, has been an advantageous model to work in this field.\\ 

Although $QCD_2$ has been widely studied, for instance, about confinement\cite{Callan:1975ps}, scattering\cite{Brower:1977hx,Ellis:2003mi,Frishman:2002ng}, and heavy quarks\cite{Bigi:1998kc}, it has been recently shown that the Lorentz theories in two dimensions admit new terms as we found in Very Special Relativity (VSR)\cite{Alfaro:2019snr}. Originally, VSR was conceived in four dimensions by Cohen and Glashow to explain the neutrino mass neither by invoking new particles nor leptonic number violation\cite{Cohen:2006ky,Cohen:2006ir}. In this theory, the main feature is the null vector $n=(1,0,0,1)$, which transforms with a phase under the SIM(2) transformations. It allows us to introduce new terms like $n\cdot p/n\cdot k$ in the lagrangian. These theories have been studied in four dimensions in electrodynamics\cite{Cheon:2009zx}, the electroweak sector\cite{Alfaro:2015fha}, non-commutative framework\cite{SheikhJabbari:2008nc}  and as a background theory\cite{Ilderton:2016rqk}.\\

In two dimensions, since the Lorentz Group is a one-parameter group; hence, there are not continuous subgroups, we cannot define a $SIM$ subgroup as in VSR theories. However, in the previous work \cite{Alfaro:2019snr} the null vector $n=(1,1)$ was shown to transform with a phase under Lorentz transformations. We will review in more detail this idea in Section II for the sake of self-completeness of this work. Thus, the possibility to add new terms to the two-dimensional models allows us to see new possible effects. In this work, we will consider this new two dimensional Lorentz invariant terms to compute the meson spectrum, following the work of 't Hooft.\\

The outline of this work is as follows. In Section II, we review the Lorentz group and the possibility to add VSR terms. In Section III, we describe the $QCD_2$ model with the new terms, and we establish the notation to work in the Light Cone Coordinates (LCC) and the Light Cone Gauge (LCG). In Section IV, we compute the quark Self-Energy and the ladder diagram that contributes to the meson spectrum using a mass for the gluon to regulate the infrared divergences. Here, we compute the full solution to the modified 't Hooft equation due to the new parameter added. Section V, is devoted to the numerical solution of this generalized equation, and finally, in section VI, we present the conclusions.

\section{Two dimensional Lorentz Group and VSR}
The full Lorentz Group for any dimension is the group of transformations that leave invariant the Minkowski metric
\be
\label{deflor}
g=\Lambda g \Lambda^T.
\ee
We use $g=diag(+1,-1)$. From this definition is clear $det(\Lambda)=\pm1$. However, we always refer to the proper Lorentz Group, as the set whose elements has determinant  $\Lambda=+1$. Now we will focus our attention on the two-dimensional case. We can represent these transformations with $2\times2$ matrices. The two dimensional proper Lorentz transformations can be parameterized by

\be
\label{transf}
\Lambda(\theta)=\begin{pmatrix}
    \cosh\theta & \sinh\theta \\
   \sinh\theta & \cosh\theta
  \end{pmatrix}.
\ee

This transformation does not have invariant tensors. However, the null vector
\be
n=\begin{pmatrix}
    1 \\
    1
  \end{pmatrix},
\ee 
transforms with a phase under this transformation, $\Lambda n=e^\theta n$. This fact means that we can add elements VSR-like to the lagrangian. For instance, we can have an additional term in the fermionic construction as $\bar{\psi} \left(\frac{i}{2} m^2 \slashed{n}(n \cdot\partial)^{-1} \right) \psi$, which was initially introduced by Cohen and Glashow to explain the neutrino mass\cite{Cohen:2006ky,Cohen:2006ir}.\\

Besides, we could add VSR mass terms for the gauge fields, as it has been studied in \cite{Alfaro:2013uva,Alfaro:2015fha,Alfaro:2019koq} whose form is $- \frac{m_g^2}{2} (n^{\alpha} F_{\mu \alpha})(n \cdot D)^{-2} (n_{\beta} F^{\mu \beta})$, where $F_{\mu\nu}$ is the field strength for a gauge field $A(x)$ and $D_\mu$ is the covariant derivative. This possibility is very interesting, because this term is gauge invariant and although in higher dimensions is not Lorentz invariant, here it is. 

\section{$QCD_2$ in the Light Cone Coordinates}

We start with the lagrangian for $QCD_2$
\be
\label{lagqcd2}
\L=- \frac{1}{4} F^a_{\mu \nu} F^{a \mu \nu} - \frac{m^2_g}{2} (n^{\alpha} F^a_{\mu \alpha}) \frac{1}{(n \cdot D)^2}
(n_{\beta} F^{a \mu \beta})+ \bar{q}^a \left( i \slashed{D} - M_a
+ i \frac{m^2}{2} \frac{\slashed{n}}{n \cdot D} \right) q^a,
\ee
where $F^a_{\mu\nu}=\partial_\mu A^a_\nu-\partial_\nu A^a_\mu+g f^{abc}A^b_\mu A^c_\nu$, with $f^{abc}$ the structure constant of the local gauge group. In addition, $D_\mu=\partial_\mu-i g A_\mu$, where $A_\mu=A^a_\mu\tau^a$, with $\tau^a$ the generators of the gauge symmetry group. Moreover, $M_a$ is the Lorentz invariant mass for the quark $q^a$. The term $m$ is a VSR mass term, which in four dimensional $SIM(2)$ theories we associate with the neutrino mass. The term $m_g$ is the gluon mass. We notice this lagrangian is gauge invariant under the transformation $A^a_\mu\to A'^a_\mu=A^a_\mu+\frac{1}{g}\partial_\mu\chi^a+f^{abc}A^b_\mu\chi^c$ and when $m=m_g=0$ we recover the standard $QCD_2$.\\

In order to manage the non abelian field strength and the non-local terms, we will write the lagrangian (\ref{lagqcd2}) in the Light Cone Coordinates (LCC). We define
\bea
x^+ = \frac{1}{\sqrt{2}} (x^0 + x^1),\\
x^- = \frac{1}{\sqrt{2}} (x^0 - x^1),
\eea

In these coordinates the metric components are $g_{++}=g_{--}=0$ and $g_{+-}=g_{-+}=1$. Thus, $x^+=x_{-}$ and $x^-=x_+$. In addition, for any vectors $R$ and $S$
\be
R\cdot S=g_{\mu\nu}R^\mu S^\nu=R^+S^-+R^-S^+.
\ee

Furthermore, for contractions with gamma matrices
\be
\slashed{R}=\gamma^{\mu} R_{\mu} = g_{\mu \nu} \gamma^{\nu} R^{\mu} = (\gamma^- R^+ + \gamma^+ R^-),
\ee
where $\gamma^{\pm} = \frac{1}{\sqrt{2}} (\gamma^0 \pm \gamma^1)$, and they satisfy $\{ \gamma^+, \gamma^- \} = 2 I$ and $(\gamma^+)^2 = (\gamma^-)^2 = 0$. Choosing $\gamma^0 = \sigma_1$ and $\gamma^1 = i \sigma_2$ we have the following matrix representation

\be
\gamma^+ = \left(\begin{array}{cc}
  0 & \sqrt{2}\\
  0 & 0
\end{array}\right), \qquad \gamma^- = \left(\begin{array}{cc}
  0 & 0\\
  \sqrt{2} & 0
\end{array}\right).
\ee

We take our null vector $n$, where $n^0=1$ and $n^1=1$. Its components in the LCC are $n^+ = \sqrt{2}$ and $n^- = 0$. Thus, for any vector $R$, the inner product reads $n\cdot R=\sqrt{2}R^-$.\\

We will work in the Light Cone Gauge (LCG), here $n\cdot A=0$. In the LCC it means $A^-=A_+=0$. Thus, $F^a_{+ -} = \partial_+ A_-^a$. Hence, $F^a_{\mu \nu} F^{a \mu \nu} =  - 2 (\partial_+ A_-^a)^2$. Moreover, we notice the term with $m_g^2$ in these coordinates is zero. Thus, we can write the equation (\ref{lagqcd2}) as
\be
\label{lagqcd2lcc}
\L=- \frac{1}{2} A_-^a(\partial_+)^2 A_-^a + \bar{q}^a \left( i
\slashed{\partial} + g (\gamma^- A_-^a) - M_a + i \frac{m^2}{2}
\frac{\gamma^-}{\partial_+} \right) q^a.
\ee

With this gauge and coordinate transformation we have only the true degrees of freedom. In this gauge the Fadeev-Popov ghost decouples from the S-matrix. Thus, only $A_-$ contributes to the S-matrix. In addition, the gluon self-interaction vanishes in the LCG, making easier the computations. The Feynman rules are listed in table \ref{feynman}, where we have defined $M_{e a}^2=M_a^2+m^2$.\\

\begin{table}[h!]
\label{feynman}
\begin{tabular}{|l|l|}
\hline
 $A_-$ propagator &  $\frac{i}{(k^-)^2}$ \\ \hline
$q^a$ propagator & $\frac{i \left( \slashed{p} + M_a - \frac{m^2}{2} \frac{\gamma^-}{p^-}
\right)}{p^2 - M^2_{e a}-i\epsilon}$ \\ \hline
Vertex & $i g \gamma^-$ \\ \hline
\end{tabular}
\caption{Table with the Feynman rules for the lagrangian in the equation (\ref{lagqcd2lcc}).}
\end{table}

We notice that the new term in the quark propagator contains $\gamma^-$. Since the vertex has the same gamma matrix, this new term will not contribute to the computations. Even so, although this part is not relevant, in the denominator of the propagator the mass is no longer the standard Lorentz invariant mass squared, but it is corrected adding $m^2$. However, this modification is purely a mass shift, and it does not mean a physical difference. In this way, the terms containing the vector $n$ does not make a difference from the standard computation.

\section{Quark Self-Energy and Ladder Diagram computation}
Since the VSR term in the quark propagator does not contribute in our computations, we will have the same expression for the Self-Energy given by 't Hooft\cite{tHooft:1974pnl} with the difference in the mass $M_{e a}^2$. Although there is the same computation we proceed with a different method. We will add a mass term for the gluon as a regulator. At the end of the computation we will make $m_g\to0$. Thus,
\be
\Sigma (p^-) = \frac{g^2}{(2 \pi)} \int_{-\infty}^{\infty} d k^- \frac{1}{(
k^-)^2 + m^2_g} sgn (k^- + p^-),
\ee
where $\Sigma$ is the sum of the irreducible self-energy parts after having eliminated the $\gamma$ matrices following the 't Hooft work.\\

We use the definition of the $sgn$ function and
\be
\label{sigmap}
\Sigma (p^-) = \frac{g^2}{(2 \pi)} \underset{\Lambda \rightarrow
\infty}{\lim} \left[ \int^{\Lambda}_{- p^-} d k^- \frac{1}{( k^-)^2
+ m^2_g} - \int^{- p^-}_{- \Lambda} d k^- \frac{1}{(
k^-)^2 + m^2_g} \right].
\ee
We compute the indefinite integral
\be
\int d k^- \frac{1}{(k^-)^2 + m^2_g} = \frac{1}{m_g}\arctan{\left( \frac{k^-}{m_g} \right)},
\ee
and evaluating in (\ref{sigmap}) we get the result for the quark Self-Energy

\be
\label{selfenergy}
\Sigma (p^-) = \frac{g^2}{\pi m_g} \arctan{\left( \frac{p^-}{m_g} \right)}.
\ee

To recover the standard result we can expand in series the $\arctan$ for $m_g\approx0$
\be
\arctan{\left( \frac{p^-}{m_g} \right)}\approx\left(\frac{\pi}{2} sgn(p^-) - \frac{m_g}{p^-}\right).
\ee
Therefore,
\be
\Sigma (p^-) \approx \frac{g^2}{\pi} \left( \frac{\pi}{2 m_g} sgn(p^-) - \frac{1}{p^-}\right).
\ee
If we define $\lambda=\frac{2m_g}{\pi}$, we recover the result of 't Hooft. The result in (\ref{selfenergy}) contains the standard result, and here the gluon mass $m_g$ acts as the infrared regulator instead of the use of the $\lambda$ cutoff in the integration over $k^-$.\\

\begin{figure}[h!]
\centering
\includegraphics[scale=0.4]{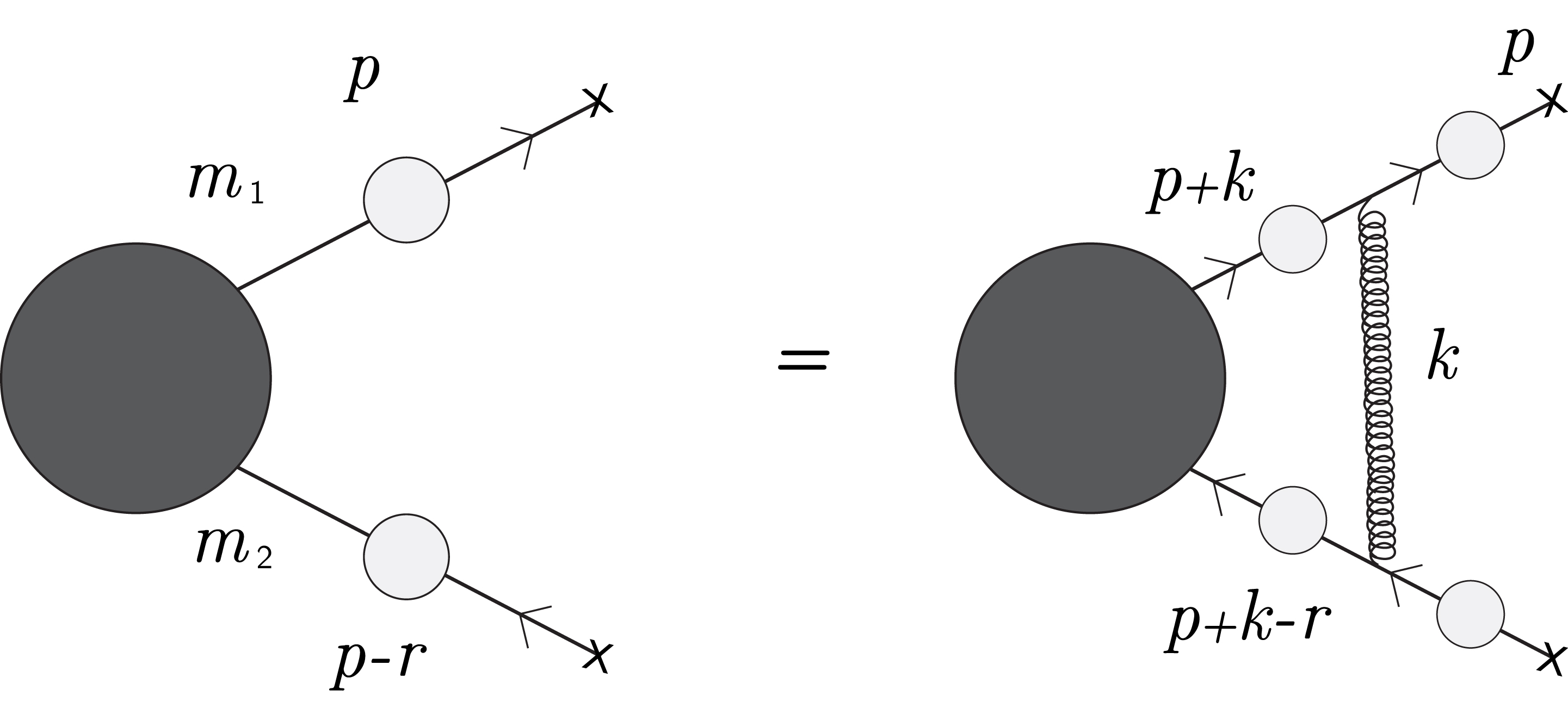}
\caption{Bethe-Salpeter equation in diagram.}
\label{fig:blob}
\end{figure}

Now, we compute the ladder diagram, which satisfies the Bethe-Salpeter equation with a quark with mass $M_1$ and momentum $p$, and an antiquark with mass $M_2$ and momentum $r-p$. Therefore,

\bea
\label{psi}
\psi (p, r) &=& \frac{4 i g^2}{(2 \pi)^2} (p^- - r^-) p^- \left( \frac{1}{2 p^+
p^- - M^2_{e 1} - \frac{g^2}{\pi m_g} \arctan{\left( \frac{p^-}{m_g}
\right)} p^- - i \varepsilon} \right) \times\nn\\
& &\times\left( \frac{1}{2 (p^+ - r^+) (p^- - r^-)
- M^2_{e 2} - \frac{g^2}{\pi m_g} \arctan{\left( \frac{p^- - r^-}{m_g}
\right)} (p^- - r^-) - i \varepsilon} \right) \int d k^+ d k^- \frac{\psi (p +
k, r)}{(k^-)^2 + m^2_g},
\eea
where $\psi$ is the blob out of which comes the quark and antiquark, as is depicted in figure \ref{fig:blob}. Moreover, we have used the dressed propagator
\be
\frac{i p^-}{p^2 - M^2_{e a} - \Sigma (p^-) p^- - i
\varepsilon}=\frac{i p^-}{2 p^+ p^- - M^2_{e a} - \frac{g^2}{\pi m_g}
\arctan{\left( \frac{p^-}{m_g} \right)} p^- - i \varepsilon},
\ee
after eliminate the gamma matrices as before.\\

Writing $\varphi (p^-, r) = \int \psi (p^+, p^-, r) d p^+$ in (\ref{psi}) we have
\bea
\label{phi}
\varphi (p^-, r) &=& \frac{i g^2}{(2 \pi)^2} \int d p^+ \left( \frac{1}{p^+ -
\frac{M^2_{e 1}}{2 p^-} - \frac{g^2}{2 \pi m_g} \arctan{\left(
\frac{p^-}{m_g} \right)} - i \varepsilon sgn(p^-)} \right)\times\nn\\
& &\times \left(
\frac{1}{p^+ - r^+ - \frac{M^2_{e 2}}{2 (p^- - r^-)} - \frac{g^2}{2 \pi m_g}
\arctan{\left( \frac{p^- - r^-}{m_g} \right)} - i \varepsilon sgn(p^- - r^-)} \right) \int d k^- \frac{\varphi (p^- + k^-, r)}{(k^-)^2 +
m^2_g}.
\eea
We notice the integration in $p^+$ is non zero if $sgn(p^- - r^-) = - sgn(p^-)$. If we choose $r^->0$ the only interval when this relation is satisfied is when $0<p^-<r^-$. Thus,
\be
\label{phi2}
\varphi (p^-, r) = - \frac{g^2}{2 \pi} \left( \frac{1}{r^+ + \frac{M^2_{e
2}}{2 (p^- - r^-)} - \frac{M^2_{e 1}}{2 p^-} + \frac{g^2}{2 \pi m_g} \left[
\arctan{\left( \frac{p^- - r^-}{m_g} \right)} - \arctan{\left(
\frac{p^-}{m_g} \right)} \right] - i \varepsilon} \right) \int^{r^- - p^-}_{-
p^-} d k^- \frac{\varphi (p^- + k^-, r)}{(k^-)^2 + m^2_g}.
\ee
We define $\alpha_i = \frac{\pi M^2_{e i}}{g^2} - 1$, $\frac{p^-}{r^-} = x$, $\frac{2 \pi}{g^2} r^+ r^- = \mu^2$ and $\mu_g = \frac{m_g}{r^-}$. Therefore
\be
\label{maineq}
- \varphi (x) \left( \mu^2 - \left( \frac{\alpha_1 + 1}{x} + \frac{\alpha_2 +
1}{(1 - x)} \right) - \frac{1}{\mu_g} \left[ \arctan{\left( \frac{1 -
x}{\mu_g} \right)} + \arctan{\left( \frac{x}{\mu_g} \right)} \right]
\right) = \int^1_0 d y \frac{\varphi (y)}{(y - x)^2 + \mu^2_g}.
\ee

We notice the gluon mass term modifies the 't Hooft result, which is contained in the limit $\mu_g\to0$. First, we notice the $\arctan$ will give a divergent part in this limit,
\be
- \varphi (x) \left( \mu^2 - \left( \frac{\alpha_1}{x} + \frac{\alpha_2}{(1 -
x)} \right) - \frac{\pi}{\mu_g} \right) = \int^1_0 d y \frac{\varphi (y)}{(y -
x)^2}.
\ee
Nevertheless, the integral in the right hand side is divergent when $y=x$. If we come back to the equation (\ref{phi}) and we use the limit $m_g\to0$; we can split the integral in the right hand side as
\be
\int d k^- \frac{\varphi (p^- + k^-, r)}{(k^-)^2} = \int^{- \lambda}_{-
\infty} d k^- \frac{\varphi (p^- + k^-, r)}{(k^-)^2} + \int^{\lambda}_{-
\lambda} d k^- \frac{\varphi (p^- + k^-, r)}{(k^-)^2} +
\int^{\infty}_{\lambda} d k^- \frac{\varphi (p^- + k^-, r)}{(k^-)^2}.
\ee
Thus, taking the principal value,
\be
P\int d k^- \frac{\varphi (p^- + k^-, r)}{(k^-)^2} = P \int d k^-
\frac{\varphi (p^- + k^-, r)}{(k^-)^2} + \frac{2}{\lambda} \varphi (p^-, r).
\ee
If we choose $\lambda=\frac{2m_g}{\pi}$ as in the quark Self-Energy we have
\be
P\int d k^- \frac{\varphi (p^- + k^-, r)}{(k^-)^2} = P \int d k^-
\frac{\varphi (p^- + k^-, r)}{(k^-)^2} + \frac{\pi}{\mu_g} \varphi (p^-, r).
\ee
We notice this last term cancels the divergent part of the $\arctan$. So, using a change of variables, we put all together and we will get
\be
\label{thoofteq}
- \varphi (x) \left( \mu^2 - \left( \frac{\alpha_1}{x} + \frac{\alpha_2}{(1 -
x)} \right) \right) = P \int^1_0 d y \frac{\varphi (y)}{(y - x)^2},
\ee
which is the t'Hooft result. Thus, the equation (\ref{thoofteq}) is a limit case of (\ref{maineq}) when $\mu_g\to0$.\\

Despite the physical content is given by the equation (\ref{thoofteq}), we will solve the generalized equation (\ref{maineq}) in order to see deviations from the standard result and the behavior of the introduced parameter. However, the equation (\ref{maineq}) is not solvable analytically. To solve it, we will proceed similarly to the 't Hooft work. First, we can write the equation (\ref{maineq}) as
\be
\label{hamiltonian}
H(x)\varphi(x)=\mu^2\varphi(x),
\ee
where
\be
H (x) = \left( \frac{\alpha_1 + 1}{x} + \frac{\alpha_2 + 1}{(1 - x)} \right)
+ \frac{1}{\mu_g} \left[ \arctan{\left( \frac{1 - x}{\mu_g} \right)} +
\arctan{\left( \frac{x}{\mu_g} \right)} \right] - \int^1_0 d y
\frac{1}{(y - x)^2 + \mu^2_g}.
\ee

From (\ref{hamiltonian}) it is explicit that the values for the meson mass squared $\mu^2$ are the eigenvalues of the operator $H$. We associate this operator with a kind of Hamiltonian. It is easy to see it, considering the functional 
\be
S=\langle \varphi, H\varphi \rangle-\mu^2\langle \varphi, \varphi \rangle,
\ee
where the inner product is defined as
\be
\label{inner}
\langle\psi(x),\phi(x)\rangle=\int_0^1 d x \psi^{*}(x)\phi(x).
\ee
When we minimize the functional $S$, we found the equation (\ref{maineq}). Moreover, we observe $H(x)$ is hermitian since we can take the following inner product
\bea
\langle \psi, H\varphi \rangle &=& \int^1_0 d x \left( \frac{\alpha_1 + 1}{x}
+ \frac{\alpha_2 + 1}{(1 - x)} \right) \psi^{\ast} (x) \varphi (x) +
\frac{1}{\mu_g} \int^1_0 d x \left[ \arctan{\left( \frac{1 - x}{\mu_g}
\right)} + \arctan{\left( \frac{x}{\mu_g} \right)} \right] \psi^{\ast} (x)
\varphi (x) \nn\\
& &- \int^1_0 d x \int^1_0 d y \frac{\psi^{\ast} (x) \varphi (y)}{(y
- x)^2 + \mu^2_g},
\eea
and it is explicit $\langle \psi, H\varphi \rangle=\langle H\psi, \varphi \rangle$ for any $\psi$ and $\varphi$ continuous in $[0,1]$. In order to avoid divergences when $x=0$ and $x=1$ we require $\varphi(0)=\varphi(1)=0$. 
Thus, we have settled our boundary condition. Hence, we will observe the behavior in $x=0$. Similarly to the 't Hooft work, we will use the ansatz $\varphi(x)\sim x^\beta$ when $x\to0$. In this way
\be
\label{cond}
(\alpha_1 + 1) x^{\beta - 1} = \int^1_0 d y \frac{y^{\beta - 1}}{y^2 +\mu^2_g}.
\ee
In this case, the only option is $\beta=1$, since the integral on the right hand side of equation (\ref{cond}) is finite. Therefore, for $\mu_g\neq0$, we have $\varphi(x)\sim x$ when $x\to0$. The same situation stands for $x=1$.\\

Before we compute the numerical solution of the integral equation, we wish to study the behavior of the eigenvalues $\mu^2$ for limit cases of $\mu_g$. First, we will consider $\mu_g$ small. In this limit, and considering $\mu^2$ large, the equation (\ref{maineq}) now is
\be
- \varphi (x) \left( \mu^2 - \frac{\pi}{\mu_g} \right) = \int^1_0 d y \frac{\varphi (y)}{(y -
x)^2+\mu^2_g},
\ee
where we took $\mu_g$ small but not zero. Since in this limit we approach to the 't Hooft solution, we use a similar ansatz $\varphi(y)=e^{i\lambda y}$, with $\lambda>0$ and $\lambda$ large. The integral in the right hand side is given by
\be
  \int^1_0 d y \frac{e^{i\lambda y}}{(y - x)^2 + \mu^2_g} \sim \int^{\infty}_{-
  \infty} d y \frac{e^{i\lambda y}}{(y - x)^2 + \mu^2_g} = 
   \frac{\pi}{\mu_g} e^{- \lambda \mu_g}
  e^{i \lambda x}.
\ee

Thus, 
\be
\mu^2 = \frac{\pi}{\mu_g} (1 - e^{- \lambda \mu_g}).
\ee

Since our boundary conditions are $\varphi(0)=\varphi(1)=0$, we must take the imaginary part of the exponential solution as $\varphi(x)=\sin{\lambda x}$. Therefore, $\lambda=n\pi$. Hence,

\be
\label{musmall}
\mu^2 = \frac{\pi}{\mu_g} (1 - e^{- n\pi \mu_g}).
\ee
Notice that:
\be
\label{musmall2}
\lim_{\mu_g\to0} \frac{\pi}{\mu_g} (1 - e^{- n\pi \mu_g})=\pi^2 n,
\ee
which is t'Hooft result.\\

On the other hand, for large $\mu_g$, the equation (\ref{maineq}) reads

\be
\varphi (x) = - \frac{1}{\mu^2_g \left( \mu^2 - \left( \frac{\alpha_1 + 1}{x}
+ \frac{\alpha_2 + 1}{(1 - x)} \right)-\frac{1}{\mu^2_g} \right)} \int^1_0 d y \varphi (y).
\ee
Integrating respect to $x$ we can get
\be
\label{muginteg}
\mu^2_g = - \int^1_0 d x \frac{1}{\mu^2 - \left( \frac{\alpha_1 + 1}{x} +
\frac{\alpha_2 + 1}{(1 - x)} \right)-\frac{1}{\mu^2_g}}.
\ee

We can neglect the term $1/\mu_g^2$ in the denominator, and we solve the integral. Since $\mu_g^2\to\infty$, the values of $\mu^2$ are those where the result of the integral goes to infinity in the right hand side for any choice of $\alpha_1$ and $\alpha_2$. We get two values

\be
\mu^2=(\sqrt{\alpha_1+1}\pm\sqrt{\alpha_2+1})^2.
\ee

We rule out the solution with a minus sign because for equal masses ($\alpha_1=\alpha_2$), when we replace this value in (\ref{muginteg}), the integral is finite. Hence, this solution does not match with $\mu_g^2\to\infty$. Therefore, the only value is given by

\be
\label{mu2large}
\mu^2=(\sqrt{\alpha_1+1}+\sqrt{\alpha_2+1})^2.
\ee

 This equation will be important in the next section to check the good approximation of the first eigenvalues.\\

\section{Numerical solution}
To compute the spectrum of equation (\ref{maineq}) for different values of the quark masses, parameterized by $\alpha_i$,  we will write $\varphi(x)$ in terms of a basis of sine functions, which have the same behavior in the boundary. Hence,
\be
\varphi (x) = \sum_{n = 1}^N A_n \sin{(n \pi x)},
\ee
and the equation (\ref{hamiltonian}) reads
\bea
\label{hamrep}
\mu^2 \sum_{m
= 1}^N A_m \sin{(m \pi x)} &=& \sum_{m = 1}^N A_m \left( \frac{\alpha_1 + 1}{x} + \frac{\alpha_2 + 1}{(1-x)} \right) \sin{(m \pi x)} + \frac{1}{\mu_g} \sum_{m = 1}^N A_m \left[
\arctan{\left( \frac{1 - x}{\mu_g} \right)} + \arctan{\left(
\frac{x}{\mu_g} \right)} \right] \sin{(m \pi x)}\nn\\
& &- \sum_{m = 1}^N A_m
\int^1_0 d y \frac{\sin{(m \pi y)}}{(y - x)^2 + \mu^2_g}.
\eea
We apply the inner product (\ref{inner}) in (\ref{hamrep}) with another basis element $\sin{(n\pi x)}$. Since the sine functions are orthogonal, we get
\bea
\mu^2 A_n &=& 2 \sum_{m = 1}^N A_m \int^1_0 d x \sin{(n \pi x)} \left( \frac{\alpha_1
+ 1}{x} + \frac{\alpha_2 + 1}{(1 - x)} \right) \sin{(m \pi x)} \nn\\
& &+\frac{2}{\mu_g} \sum_{m = 1}^N A_m \int^1_0 d x \sin{(n \pi x)} \left[
\arctan{\left( \frac{1 - x}{\mu_g} \right)} + \arctan{\left(
\frac{x}{\mu_g} \right)} \right] \sin{(m \pi x)}\nn\\
& &- 2 \sum_{m = 1}^N A_m
\int^1_0 d x \sin{(n \pi x)} \int^1_0 d y \frac{\sin{(m \pi y)}}{(y
- x)^2 + \mu^2_g}.
\eea

\begin{figure}[h!]
\centering
\includegraphics[scale=0.58]{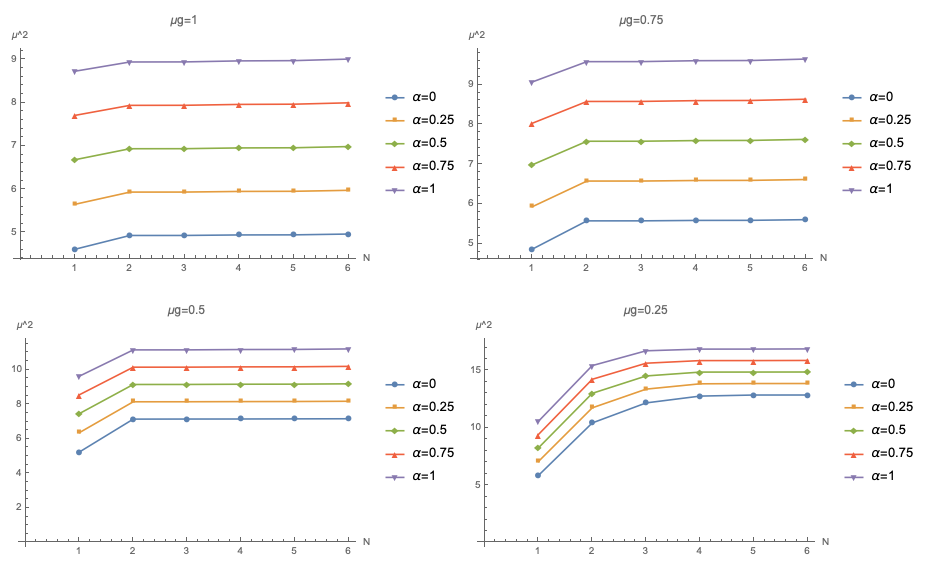}
\caption{Plot for the different values of $\alpha$ for every value of $\mu_g$ chosen.}
\label{fig:eigen}
\end{figure}

\begin{figure}[h!]
\centering
\includegraphics[scale=0.6]{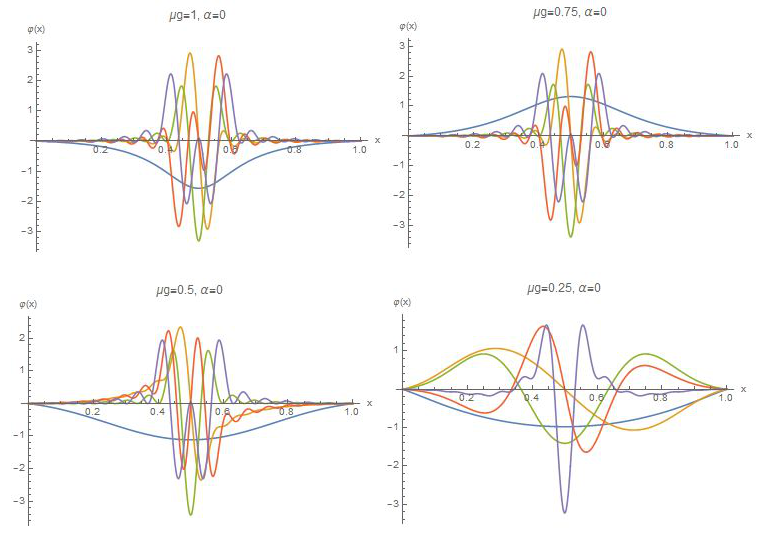}
\caption{Plot of $\varphi(x)$ for the first five eigenvalues using $N=30$ considering $\alpha=0$ and the different values of $\mu_g$.}
\label{fig:eigenfun}
\end{figure}

To write it in a compact form, we call $H_{n m} = 2 \int^1_0 d x \sin{(n \pi x)} H (x) \sin{(m \pi x)}$. Therefore,
\be
\sum_{m = 1}^N A_m H_{n m} = \mu^2 A_n.
\ee
This equation is a system of equations that we could represent in a matricial way as
\be
\left(\begin{array}{cccc}
  H_{11} - \mu^2 & H_{12} & \ldots & H_{1 N}\\
  H_{21} & H_{22} - \mu^2 & \ldots & H_{2 N}\\
  \vdots & \ddots &  & \vdots\\
  H_{N 1} & H_{N 2} & \cdots & H_{N N} - \mu^2
\end{array}\right) \left(\begin{array}{c}
  A_1\\
  A_2\\
  \vdots\\
  A_N
\end{array}\right) = 0.
\ee
We solve this system using a computer, considering $\alpha_1=\alpha_2=\alpha$, which means a meson built from equal mass quarks. We have taken the series up to $N=60$. To check the precision in the estimation, first we run the code using $\mu_g=10^{20}$ for $\alpha=0$ and $\alpha=1.2$. We got a relative error between the sixth eigenvalue obtained from the numerical computation and the value obtained using the equation (\ref{mu2large}) of $0.97\%$. It means the code gives with good accuracy the first eigenvalues. To show some examples we use the values for $\mu_g=\{0.25,0.5,0.75,1\}$ and for every one we compute $\alpha=\{0,0.25,0.5,0.75,1\}$. The table with the first six eigenvalues for every choice of $\mu_g$ and $\alpha$ is displayed in Table \ref{tab:mug}, and figure \ref{fig:eigen} shows the plots of $\mu^2$ as function of $N$. Also, we have plotted $\varphi(x)$ corresponding to the first five eigenvalues for $\mu_g=\{0.25,0.5,0.75,1\}$ and $\alpha=0$ using $N=30$. We show in figure \ref{fig:eigenfun} those plots.\\

\begin{table}[h!]
\begin{tabular}{|l|l|l|l|l|}
\hline
\multicolumn{5}{|c|}{$\mu_g=1$}                          \\ \hline
$\alpha=0$       & $\alpha=0.25$     & $\alpha=0.5$      & $\alpha=0.75$     & $\alpha=1$        \\ \hline
4.61439552 & 5.65176658 & 6.68077709 & 7.70404365 & 8.7231667 \\ \hline
4.93136662 & 5.93243153 & 6.93349634 & 7.93456109 & 8.93562581 \\ \hline
4.93189351 & 5.93317101 & 6.9344831  & 7.93583031 & 8.93721311\\ \hline
4.9436361 & 5.94790939 & 6.95218227& 7.95645494 & 8.96072748 \\ \hline
4.94573975 & 5.95084671 & 6.95607719 & 7.96142777 & 8.96689414 \\ \hline
4.96427122 & 5.9739377 & 6.98360333 & 7.9932685 & 9.00293338 \\ \hline
\end{tabular}
\\
\begin{tabular}{|l|l|l|l|l|}
\hline
\multicolumn{5}{|c|}{$\mu_g=0.75$}                          \\ \hline
$\alpha=0$       & $\alpha=0.25$     & $\alpha=0.5$      & $\alpha=0.75$     & $\alpha=1$        \\ \hline
4.85904093 & 5.92629341 & 6.98059157 & 8.02562527  & 9.06373378 \\ \hline
5.57179767  & 6.57286595 & 7.57393275 & 8.57499881 & 9.57606446 \\ \hline
5.57224874 & 6.57346191 & 7.57469377  & 8.57594455 & 9.57721446 \\ \hline
5.58322844 & 6.58751459& 7.5917951  & 8.59607281 & 9.60034893 \\ \hline
5.58503682 & 6.58990219 & 7.59483931 & 8.59984789 & 9.6049274 \\ \hline
5.60247619 & 6.61216952 & 7.62185117 & 8.63152696  & 9.64119941  \\ \hline
\end{tabular}
\\
\begin{tabular}{|l|l|l|l|l|}
\hline
\multicolumn{5}{|c|}{$\mu_g=0.5$}                          \\ \hline
$\alpha=0$       & $\alpha=0.25$     & $\alpha=0.5$      & $\alpha=0.75$     & $\alpha=1$        \\ \hline
5.23576212 & 6.35770772 & 7.46047111 & 8.54901673 & 9.62657471 \\ \hline
7.1403603 & 8.14478232 & 9.14623622 & 10.1474373 & 11.14857094  \\ \hline
7.14502363 & 8.14620122 & 9.14738569 & 10.14857774  & 11.14977767 \\ \hline
7.14789743 & 8.15515738 & 9.16047816 & 10.16518965 & 11.16969811 \\ \hline
7.15538285 & 8.1601082 & 9.16486118 & 10.16964417 & 11.17445815 \\ \hline
7.16159456 & 8.1739512 & 9.18504669 & 10.19543122 & 11.20551428 \\ \hline
\end{tabular}
\begin{tabular}{|l|l|l|l|l|}
\hline
\multicolumn{5}{|c|}{$\mu_g=0.25$}                          \\ \hline
$\alpha=0$       & $\alpha=0.25$     & $\alpha=0.5$      & $\alpha=0.75$     & $\alpha=1$        \\ \hline
5.84025691 & 7.05575435 & 8.24374784 & 9.41097461 & 10.56186585 \\ \hline
10.42360001 & 11.74136082  & 13.005966   & 14.23085804 & 15.42480511  \\ \hline
12.17114966 & 13.3723625 & 14.52208945 & 15.63394044 & 16.71669344 \\ \hline
12.76468806 & 13.8362085 & 14.85816306 & 15.86133482 & 16.86297696  \\ \hline
12.85839548 & 13.8601697 & 14.86167676 & 15.86325231 & 16.86499552 \\ \hline
12.85916965 & 13.86117434 & 14.86703374 & 15.87466975 & 16.88074613 \\ \hline
\end{tabular}
\caption{Table of the first six eigenvalues for every $\mu_g$ and $\alpha$.}
\label{tab:mug}
\end{table}

\section{Conclusions}

We have analyzed the effects of adding new Lorentz invariant terms in the 't Hooft model for QCD2. Those terms come from the fact that the null vector $n=(1,1)$ transforms up to a phase under the Lorentz transformations. These new terms do not modify anything, except a different mass for the quark. Although there is not a difference in two dimensions, the VSR term for the gluon in higher dimensions is different than zero. Since this term has been associated  to the gluon mass\cite{Alfaro:2013uva,Alfaro:2015fha}, it could be interesting for future work to see the effects of this term in the four dimensional case. It is important to keep in mind this possibility since the gluon mass bounds are not as strong as the photon mass bounds\cite{Yndurain:1995uq}, where one of the estimated bounds using cosmological arguments was $m_g<10^{-10}$ MeV.\\

We have introduced a mass for the gluon to regulate the infrared divergence in the 't Hooft model. This mass could be thought as an effective mass whose origin should be elucidate in the future works. In any case, this method works as a different regulator. The standard result is recovered at the end when we set this mass to zero. Furthermore, we have computed numerically the general solution for any value of $m_g$. We notice from the plots a non-zero value of the parameter $m_g$ does not contains a Regge trajectory.


\begin{acknowledgements}

The work of A. Soto is supported by the CONICYT-PFCHA/Doctorado Nacional/2017-21171194 and
Fondecyt 1150390. The work of J. Alfaro is partially supported by Fondecyt 1150390 and CONICYT-PIA-ACT1417.\\

\end{acknowledgements}
\vspace{-20pt}




\end{document}